\documentclass[10pt,aps,twocolumn,superscriptaddress,preprintnumbers,nofootinbib]{revtex4-1} 

\usepackage{graphicx}
\usepackage{amssymb}
\usepackage{amsmath}
\usepackage{color}
\usepackage{float}
\usepackage{slashed}
\usepackage{bbold}

\newcommand{\beq}{\begin {equation}}  
\newcommand{\eeq}{\end   {equation}} 
\newcommand{\bea}{\begin {eqnarray}} 
\newcommand{\eea}{\end   {eqnarray}}  
\newcommand{\baa}{\begin {array}   } 
\newcommand{\eaa}{\end   {array}   }     
\newcommand{\bit}{\begin {itemize} }
\newcommand{\eit}{\end   {itemize} }
\newcommand{\be }{\begin {equation}} 
\newcommand{\ee }{\end   {equation}}
\newcommand{\nn }{\nonumber        }

\newcommand{\pt}{p_T}

\newcommand{\mt}{M_{T2}}

\def\MET{{\not \!  \! E}_T}

\def\lsim{\mathrel{\rlap{\lower4pt\hbox{$\sim$}}
   \raise1pt\hbox{$<$}}}                
\def\gsim{\mathrel{\rlap{\lower4pt\hbox{$\sim$}}
   \raise1pt\hbox{$>$}}}                

\begin{document}
	
\preprint{UTTG-28-14}
\preprint{TCC-027-14}


\title{Signatures of Top Flavored Dark Matter}

\author{Can Kilic}
\author{Matthew D. Klimek}
\author{Jiang-Hao Yu}
\affiliation{Theory Group, Department of Physics and Texas Cosmology Center,
\\The University of Texas at Austin,  Austin, Texas 78712 USA}


\begin{abstract}

We study the experimental signatures of top flavored dark matter (top FDM) in direct detection searches and at the LHC. We show that for a dark matter mass above 200~GeV, top FDM can be consistent with current bounds from direct detection experiments and relic abundance constraints. We also show that next generation direct detection experiments will be able to exclude the entire perturbative parameter region for top FDM. For regions of parameter space where the flavor partners of top FDM are not readily produced, the LHC signatures of top FDM are similar to those of other models previously studied in the literature. For the case when the flavor partners are produced at the LHC, we study their impact on a search based on transverse mass variables and find that they diminish the signal significance. However, when the DM flavor partners are split in mass by less than 120--130 GeV, the LHC phenomenology becomes very distinctive through the appearance of displaced vertices. We also propose a strategy by which all parameters of the underlying model can be experimentally determined when the flavor partners can be observed.
\end{abstract}

\maketitle


\section{Introduction}

Astrophysical observations have revealed conclusively that the dominant type of matter in the universe is a particle which is not part of the Standard Model (SM), referred to as Dark Matter (DM). While the gravitational interactions of DM have been firmly established from an observational point of view, several classes of experimental searches have so far not succeeded in providing any information about the properties of DM as a particle. In the most commonly considered classes of models, DM arises as a thermal relic with weak-scale cross sections and the preferred DM mass is not too far removed from the masses of SM particles, typically in the GeV-TeV range.

The majority of models for beyond the Standard Model physics in this energy range aim to address the Higgs mass hierarchy problem by pairing SM particles with partner particles of the same gauge quantum numbers, but not necessarily of the same spin. Since DM must not carry electromagnetic or color charge, there are only a few choices in these models for who the DM particle can be partners with, and the phenomenologically most viable identities for DM are the partners of the electroweak gauge bosons, or neutral degrees of freedom in the Higgs sector.

However, one must not forget that there is an implicit assumption in all these models, namely that DM and the solution to the Higgs mass hierarchy problem are intimately connected. There is no direct evidence for this assumption to be true, which brings up the question of whether there are any other guiding principles one might adopt in constructing models of DM. If one drops the assumption that DM is tied to the solution of the hierarchy problem, other interesting approaches present themselves, such as a possible connection between all matter particles, including DM as well as the SM matter degrees of freedom.

All SM matter degrees of freedom have a property in common: they come in three generations. The origin of the SM flavor structure is unknown, and it is assumed to arise from physics at a high energy scale. If one pursues the alternative guiding principle of looking for a connection between DM and matter particles in the SM, one is then led to the possibility that DM itself may transform nontrivially under flavor symmetries. It is therefore worthwhile to consider classes of models with ``Flavored Dark Matter'' (FDM)~\cite{Cheung:2010zf,Kile:2011mn,Batell:2011tc,Kamenik:2011nb,Agrawal:2011ze,Kumar:2013hfa,Chang:2013oia,Kile:2013ola,Bai:2013iqa,Batell:2013zwa,Agrawal:2014ufa,Agrawal:2014una,Gomez:2014lva,Agrawal:2014aoa,Hamze:2014wca,Lee:2014rba,Kile:2014jea,Blanke:2014mea}, to look for universal features of such models, and to study whether there exist experimental signatures that may be used to distinguish FDM from more traditional models of DM, such as a neutralino in a supersymmetric setup. It is even possible that this line of inquiry may yield new insights in searching for the origin of the SM flavor structure.

The general consequences of this approach and the classes of models that arise from this starting point were considered in Ref.~\cite{Agrawal:2011ze}. In particular, the FDM models of phenomenological interest contain a local interaction that couples the DM flavor multiplet to a SM matter field via a mediator, which in the most minimal setup is a flavor singlet. We will refer to this as the FDM interaction, and we will use $\phi$ to denote the mediator field. The FDM interaction has several consequences that determine the general phenomenological features of this class of models. First of all, it is natural for the DM flavors to be split in mass, because even if the DM flavors have the same mass at a high energy scale, mass splittings will be induced at loop level from the FDM interaction. Second, the heavier DM flavors can decay to the lightest one, therefore only the lightest member of the DM flavor multiplet is cosmologically stable\footnote{Unless the lifetime of the heavier flavors is much longer than the age of the universe, which is not generic and requires a rather special setup. In this paper we will only concern ourselves with the generic case.}. From this point on we will refer to the stable component of the DM multiplet as $\chi$, and to the heavier flavors collectively as $\chi'$. Third, when $\phi$ is a flavor singlet, constraints from flavor violating SM processes strongly prefer the DM and SM flavor structures to be aligned, which means that the DM multiplet should have three flavors. These then become flavor partners with the SM matter field that the FDM interaction couples them to. The flavor alignment is most easily formulated in the language of Minimal Flavor Violation, or MFV~\cite{D'Ambrosio:2002ex}, although a more general scenario has also been studied~\cite{Agrawal:2014aoa}.

At this point one can still choose $SU(2)\times U(1)$ quantum numbers for the DM multiplet as well as which SM matter multiplet it is coupled to in the FDM interaction. In the interest of minimality, we will restrict ourselves to the case where the DM multiplet is a SM gauge singlet. The SM multiplet appearing in the FDM interaction and the specific flavor in that multiplet that couples to the lightest FDM state $\chi$ is then used to label the FDM model in consideration, for example in a model of electron-FDM, $\chi$ is directly coupled to the electron in the FDM interaction. There are two possible choices even at this point, since the FDM interaction may contain the right-handed or the left-handed lepton field. Again in the interest of minimality we will choose $\phi$ to be an $SU(2)$ singlet, which means that only $SU(2)$ singlet SM fields appear in the FDM interaction, so that electron FDM would couple to the right-handed electron.

The scenario of lepton-flavored DM has been studied in the literature~\cite{Agrawal:2011ze,Hamze:2014wca,Lee:2014rba,Kile:2014jea} and shown to have many interesting features, including the possibility of having an asymmetry between the DM and its antiparticle, generated during high-scale leptogenesis. For quark flavored DM, direct detection bounds can be severe if $\chi$ is coupled to the light quarks, because tree-level $\phi$ exchange leads to a large cross section for $\chi$ scattering off of nuclei~\cite{Agrawal:2011ze}. On the other hand, when $\chi$ couples to a third generation quark, the dominant process for $\chi$-nucleus scattering arises at loop level, and direct detection bounds are less constraining. In this paper, we will turn our attention to the case of top FDM, and we will study the current exclusion bounds as well as future discovery prospects in direct detection experiments as well as in LHC searches. We will show that relic abundance considerations combined with direct detection constraints can generically be satisfied for $m_{\chi}\gsim200~$GeV. In this parameter range, not only do the existing LHC searches at 7 and 8 TeV have little exclusion power, but we will show that even at the 14 TeV LHC, searches based on traditional kinematic observables only become sensitive to top FDM at high luminosity $\gsim100~$fb$^{-1}$. When $\chi'$ is heavy and is not readily produced on shell at the LHC, the collider phenomenology becomes similar to other models that have been previously studied in the literature, in particular Ref.~\cite{Gomez:2014lva}. On the other hand, we will show that when the $\chi$ and $\chi'$ masses are split by less than 120--130~GeV, top FDM models give rise to displaced vertices that would not only make discovery significantly easier, but also allow for all parameters of the underlying model to be experimentally determined. We will also show that the next generation of direct detection experiments will be able to probe the entire perturbative parameter region that is not currently excluded.

The layout of the paper is as follows: In Sec.~\ref{sec:model} we will review the theoretical aspects of the top FDM model in detail. In Sec.~\ref{sec:DD} we will calculate the cross section for $\chi$ to scatter off of nuclei, and we will determine the region of parameter space that has been excluded by existing direct detection constraints, as well as the region that can be probed with the next generation of direct detection experiments. In Sec.~\ref{sec:collider} we will discuss the LHC signatures of top FDM, we will evaluate the prospects for a search strategy based on kinematic observables and we will study under which conditions signal events can be expected to contain displaced vertices. We will also comment on how the parameters of the model can be measured post-discovery. We will conclude in Sec.~\ref{sec:conclusions}.


\section{The Model} 
\label{sec:model}

In this section we will briefly review the theoretical features of top FDM. We will use the conventions of Ref.~\cite{Agrawal:2011ze}, in particular, we will take the DM multiplet to transform in the (anti-)fundamental representation of $U(3)_{u^{c}}$. The Lagrangian contains the FDM interaction 
\bea
{\mathcal L} \supset \lambda_{ij} \chi_{i} u^c_j \; \phi \; \; + {\rm h.c.}
\label{eq:FDMterm}
\eea
Here, $\chi_{i}$ denotes the DM flavor triplet and $\phi$ is the flavor singlet mediator. As mentioned in the Introduction, the DM multiplet will be taken to be a SM gauge singlet, which means that $\phi$ carries both $SU(3)$ color as well as hypercharge. Note that there is a global $U(1)$ symmetry under which $\phi$ has charge $+1$, all three $\chi_{i}$ have charge $-1$ and all SM fields are uncharged. This symmetry can be named ``$\chi$-number'' and it ensures that the lightest state that has nonzero $\chi$-number is stable. Since $\phi$ carries color and electric charges, it is not phenomenologically viable for it to be stable, so in this paper we will only consider the case that $\phi$ is heavier than the lightest of the $\chi_{i}$.

From the interaction term of Eq.~\ref{eq:FDMterm} it is clear that one of $\phi$ and $\chi$ must be a fermion while the other is a boson. As in Ref.~\cite{Agrawal:2011ze}, we will adopt a benchmark for the rest of this paper that $\chi$ is a Dirac fermion while $\phi$ is a complex scalar. Different setups are possible, such as making $\chi$ a boson and $\phi$ a fermion, but these do not dramatically change the qualitative features of the phenomenology.

The simplest way to be consistent with flavor constraints is to adopt the MFV paradigm~\cite{D'Ambrosio:2002ex}, namely that the SM Yukawa couplings are the only source of the breaking of flavor symmetries. Clearly, this is not intended to be a complete model. Rather it is an assumption about the UV completion of the SM in which the underlying flavor symmetry of the full model is spontaneously broken by a single source, from which both the SM Yukawa couplings and the FDM coupling of Eq.~\ref{eq:FDMterm} inherit their flavor structure. While it is a very interesting problem to construct such a UV model, this is beyond the modest goals of this paper, and we will simply adopt the MFV paradigm as an assumption here.

With the MFV paradigm, and the assumption that $U(3)_{\chi}$ is identified with $U(3)_{u^{c}}$, the $\lambda_{ij}$ matrix has a flavor structure corresponding to $3\times\bar{3}$ of $U(3)_{u^{c}}$. Assuming further that the relative normalization of the spurions $Y_{u}$ and $Y_{d}$ is preserved by the UV theory, we can then write
\begin{align}
   {\lambda_{ij}} = {\left( \alpha \mathbb{1}  + \beta \; Y_{u}^{\dagger} Y_{u}\ldots +\gamma Y_{u}^{\dagger} V_{\rm CKM} Y_{d}  Y_{d}^{\dagger} V^{\dagger}_{\rm CKM} Y_{u}  \ldots \right)_{ij}} \; ,
\label{eq:lambdamatrix}
\end{align}
where $Y_{u}=$diag$(y_{u},y_{c},y_{t})$ and $Y_{d}=$diag$(y_{d},y_{s},y_{b})$. Note that in addition to the first two terms, which are the unique contributions to order $Y^{0}$ and $Y^{2}$ with the correct flavor structure, we have also included the term with coefficient $\gamma$, which is only one of several possible terms of order $Y^{4}$ that has the correct flavor structure. The reason we have singled out this particular term is that it is the leading term that cannot be diagonalized due to quark mixing in the SM. Since $y_{b}^{2}\ll y_{t}^{2}$, for most purposes this term can be neglected against the term proportional to $\beta$, but we will return to it in Sec.~\ref{sec:displaced} where naively even a small flavor-changing coupling can make a difference, if only to show that it can be safely neglected even then.

Since $y_{u,c}\ll y_{t}$, it is in principle possible for the coupling of $\chi_{t}$ to the SM to be different than those of $\chi_{u,c}$ (which can be taken as equal to each other to a very good approximation) by an order-one factor, if $\beta\sim\alpha$. The results of this paper will be essentially independent of the fact that these two values may be different, and therefore for simplicity we will consider the case $\beta\ll\alpha$ for most of this paper, in which case all three couplings have the same value, which we will denote simply by $\lambda$. In Sec.~\ref{sec:params} we will return to this point and  consider how one might attempt to separately measure the values of $\lambda_{t}$ and $\lambda_{u,c}$ at the LHC.

MFV also dictates the flavor structure of the $\chi$ mass term to have the form
\begin{align}
\left[m_{\chi}\right]_{ij} =
{\left( m_0 \mathbb{1}  + \Delta m \; Y_{u}^{\dagger} Y_{u} \right)_{ij}}\ .
\label{eq:FDMmass}
 \end{align}
Unlike the case of the coupling matrix $\lambda_{ij}$, here the presence of the leading correction proportional to $\Delta m$ has great phenomenological significance. As mentioned in the Introduction, there cannot be any relic $\chi_{u}$ or $\chi_{c}$ left over today because in that case DM can scatter off of nuclei at tree level, and this possibility is strongly constrained by direct detection experiments (see Ref.~\cite{Agrawal:2011ze} for quantitative details). Therefore $\chi_{t}$ must be the lightest state. Depending on the value of the mass gap, the decay of $\chi_{u}$ and $\chi_{c}$ to $\chi_{t}$ may proceed through on-shell or off-shell top quarks. We will study the branching ratios for this decay in more detail in Sec.~\ref{sec:collider}. Not surprisingly, the value of $\Delta m$ is quite important for collider physics, as it will affect the kinematics of signal events containing these heavier states. Note also that since $y^{2}_{u}\ll y^{2}_{c}\ll y^{2}_{t}$, the mass gap between $\chi_{u}$ and $\chi_{c}$ is much smaller than the mass gap between either one of them and $\chi_{t}$, and therefore for all purposes of this paper, $\chi_{u}$ and $\chi_{c}$ can be taken to be degenerate. Indeed, when the very slightly heavier $\chi_{u}$ state is produced, it would decay directly to $\chi_{t}$ since the available phase space for it to decay to $\chi_{c}$ is extremely small. As already mentioned in the Introduction, we will refer to $\chi_{t}$ as $\chi$ from this point on, and both $\chi_{u}$ and $\chi_{c}$ will be denoted by $\chi'$. We will only distinguish between them when not doing so can lead to confusion.

The relic abundance of top FDM is set via the usual weakly interacting massive particle (WIMP) mechanism\footnote{As proposed in Ref.~\cite{Hamze:2014wca}, the flavor structure of FDM makes it possible to generate an asymmetry for $\chi$ in the early universe. This intriguing possibility for top FDM is beyond the scope of this paper, but will be taken up in a future study.}. In principle, coannihilations between the three $\chi$ flavors may be important. However, for the reasons outlined above, it is preferable to take $\Delta m$ to be appreciable, and we will work with the simpler scenario where all $\chi'$ have already decayed before $\chi$ freezes out. As will be shown in Sec.~\ref{sec:DD}, direct detection experiments favor $m_{\chi}\gsim200~$GeV , therefore the dominant annihilation channel is $\chi\overline{\chi} \to t \bar{t}$ through $\phi$ exchange, and the annihilation cross section ($s$-wave) is given by
\bea
	&&\sigma v_{\rm{rel}} = 3\frac{\lambda^4}{32 \pi}\frac{ m_\chi^2 }{(m_\chi^2 + m_\phi^2 - m_t^2)^2} \sqrt{ 1 - \frac{m_t^2}{m_\chi^2}},
	\label{eq:relic}
\eea

While we will not pursue this possibility further in the rest of the paper, it is worth remarking that in the case when $\chi$ is lighter than the top quark, the freezeout calculation is significantly modified, both because the top quark drops out of thermal equilibrium significantly above $m_{\chi}$, and also because the annihilation has to proceed through off-shell tops, or in extreme cases, through the channel $\chi \overline{\chi} \to g g$ via a top quark loop.


\section{Direct Detection Constraints}
\label{sec:DD}

\begin{figure}[htp]
\begin{center}
\includegraphics[width=0.23\textwidth]{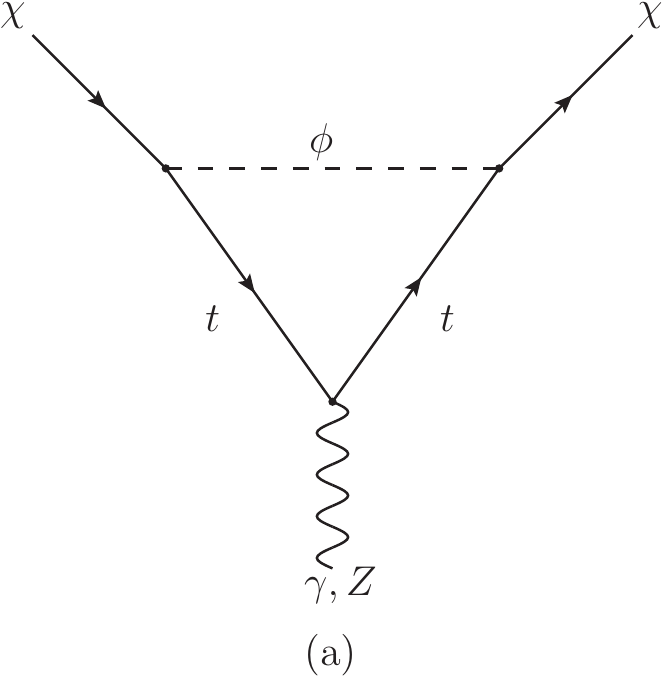} 
\includegraphics[width=0.23\textwidth]{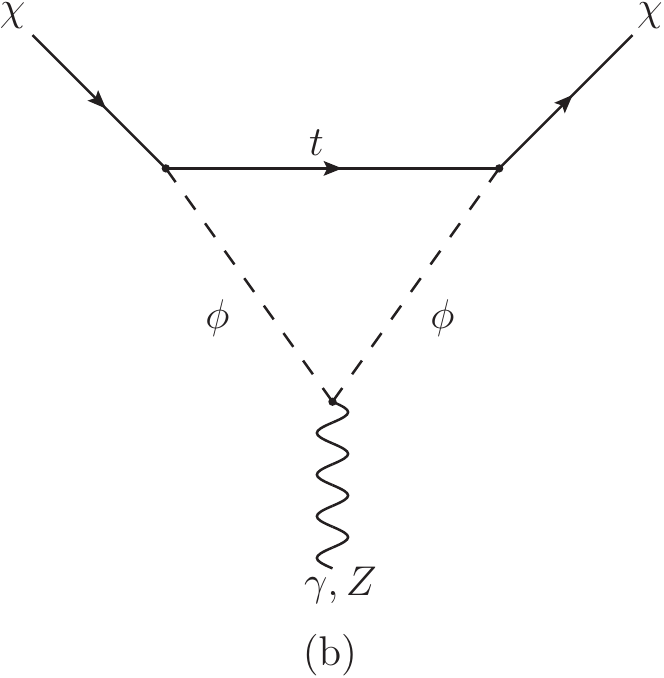} 
\caption{\small Feynman diagrams that contribute to the effective coupling to $\gamma$/$Z$ in the top FDM model. The vector boson lines in the loop diagrams can be attached to either the top quark or to the mediator $\phi$ running in the loop.}
\label{fig:vvvtri}
\end{center}
\end{figure}

The amplitude for top FDM to scatter off of a nucleus in a direct detection experiment receives contributions from a number of processes, all arising at loop level since $\chi$ is a SM singlet. In particular, at loop level $\chi$ develops charge-radius and magnetic dipole type couplings to the photon and the $Z$ boson (shown in Fig.~\ref{fig:vvvtri})\footnote{We drop magnetic dipole couplings to $Z$ since those are suppressed by $k^{2}/m_{Z}^{2}$ where $k$ is the typical momentum exchange in the scattering.}, a coupling to the Higgs boson (essentially the same process as the diagram on the left of Fig.~\ref{fig:vvvtri}, but with $\gamma$/$Z$ replaced by $h$), and effective four-point couplings to $u$-$\bar{u}$ and $g$-$g$ generated through box diagrams. These contributions were all calculated in Ref.~\cite{Kumar:2013hfa}, with the conclusion that the $\gamma$/$Z$ exchange diagram dominates over the four-point coupling to $u$-$\bar{u}$, the remaining diagrams being negligible compared to these two.

Let us start by focusing on the diagrams of Fig.~\ref{fig:vvvtri}. After integrating out $\phi$, the one-loop induced couplings of top FDM are given by the effective Lagrangian
\bea
  {\mathcal L}_{\rm eff} &=& g_\chi \bar{\chi} \gamma_\mu   \chi Z^\mu + b_\chi  \bar{\chi} \gamma_\nu   \chi \partial_\mu F^{\mu\nu} +
  \mu_\chi  \bar{\chi} i\sigma_{\mu\nu}  \chi F^{\mu\nu},\;\;\;
\label{eq:gammafdm}
\eea
where the effective coupling $g_\chi $ of the $Z$ boson, the charge radius $b_\chi$ and the magnetic dipole moment $\mu_\chi$ are given to leading order in $m_{\chi}/m_{\phi}$ by\footnote{For all plots in this paper, we use the exact expressions of $g_{\chi}$, $b_{\chi}$ and $\mu_{\chi}$ which are also valid for $m_{\chi}\lsim m_{\phi}$.}
\bea
	g_{\chi}  &=&  - \frac{3 \lambda^2 e}{32\pi^2 s_W c_W} \frac{m_t^2}{m_\phi^2} \left( 1 +  \log\frac{m_t^2}{m_\phi^2} \right) , \nn \\
	b_{\chi}  &=&  -\frac{3 \lambda^2 e Q_t}{64\pi^2 m_\phi^2} \left( 1 + \frac{2}{3}\log\frac{m_t^2}{m_\phi^2} \right) , \nn \\
	\mu_\chi  &=&  -\frac{3 \lambda^2 e Q_t m_\chi}{64\pi^2 m_\phi^2}.
\eea

Using Eq.~\ref{eq:gammafdm} we can write down an effective Lagrangian describing the scattering of $\chi$ from a light quark $q$ as
\begin{align}
	\mathcal{L}_{\rm eff} =  c^q_Z \overline{\chi} \gamma^\mu \chi \overline{q} \gamma_\mu q  + c^q_{\gamma} \overline{\chi} \gamma^\mu \chi \overline{q} \gamma_\mu q
	+ c^q_{\rm md} \overline{\chi} i\sigma^{\alpha\mu} \frac{k_\alpha}{k^2} \chi \overline{q} \gamma_\mu q,
\end{align}
where 
\bea
	c^q_Z        =  \frac{g_Z^q g_{\chi}}{m_Z^2}, \quad c^q_\gamma        =  e Q_q b_{\chi}  , \quad
	c^q_{\rm md} = e Q_q \mu_\chi,
\eea
and $g_Z^q$ is the coupling of the quark $q$ to the $Z$ boson.

We next convert the quark-level operators to nucleon-level operators and take the non-relativistic limit, arriving at the following effective Lagrangian at the nucleon level ($N=p,n$)
\bea
	\mathcal{L}_{\rm eff} &=&  c^N_Z \overline{\chi} \gamma^\mu \chi \overline{N} \gamma_\mu N +
	c^N_{\gamma} \overline{\chi} \gamma^\mu \chi \overline{N} \gamma_\mu N \nn \\
	&& + c^N_{Q} \overline{\chi}  i\sigma^{\alpha\mu}\frac{k_\alpha}{k^2} \chi \overline{N} K_\mu N \nn\\
	&& + c^N_{\mu} \overline{\chi} i\sigma^{\alpha\mu}\frac{k_\alpha}{k^2} \chi \overline{N} i\sigma^{\beta\mu} k_\beta N,
\eea
where $K_{\mu}$ is the sum of the incoming and outgoing nucleon four-momenta. Here the coefficients $c^N$ are given by
\begin{align}
	c^N_Z &= \frac{g_Z^N g_{\chi}}{m_Z^2}, \\
	c^N_\gamma &= e Q_N b_\chi,		
\end{align}
with $g^p_Z = \frac{e}{s_W c_W}(\frac12 - 2 s_W^2)$ and $g^n_Z = -\frac{e}{2 s_W c_W} $. The coefficients $c^{n}_{Q}$ and $c^{N}_{\mu}$ are given by 
\bea
	c^N_{Q} = e Q_N  \mu_\chi / 2m_{N}, \quad c^N_{\mu} = - e \tilde{\mu}_N  \mu_\chi / 2m_{N},
\eea
with $\tilde{\mu}_p = 2.8$ and $\tilde{\mu}_n = -1.9$.

\begin{figure}[tp]
\begin{center}
\includegraphics[width=0.55\textwidth]{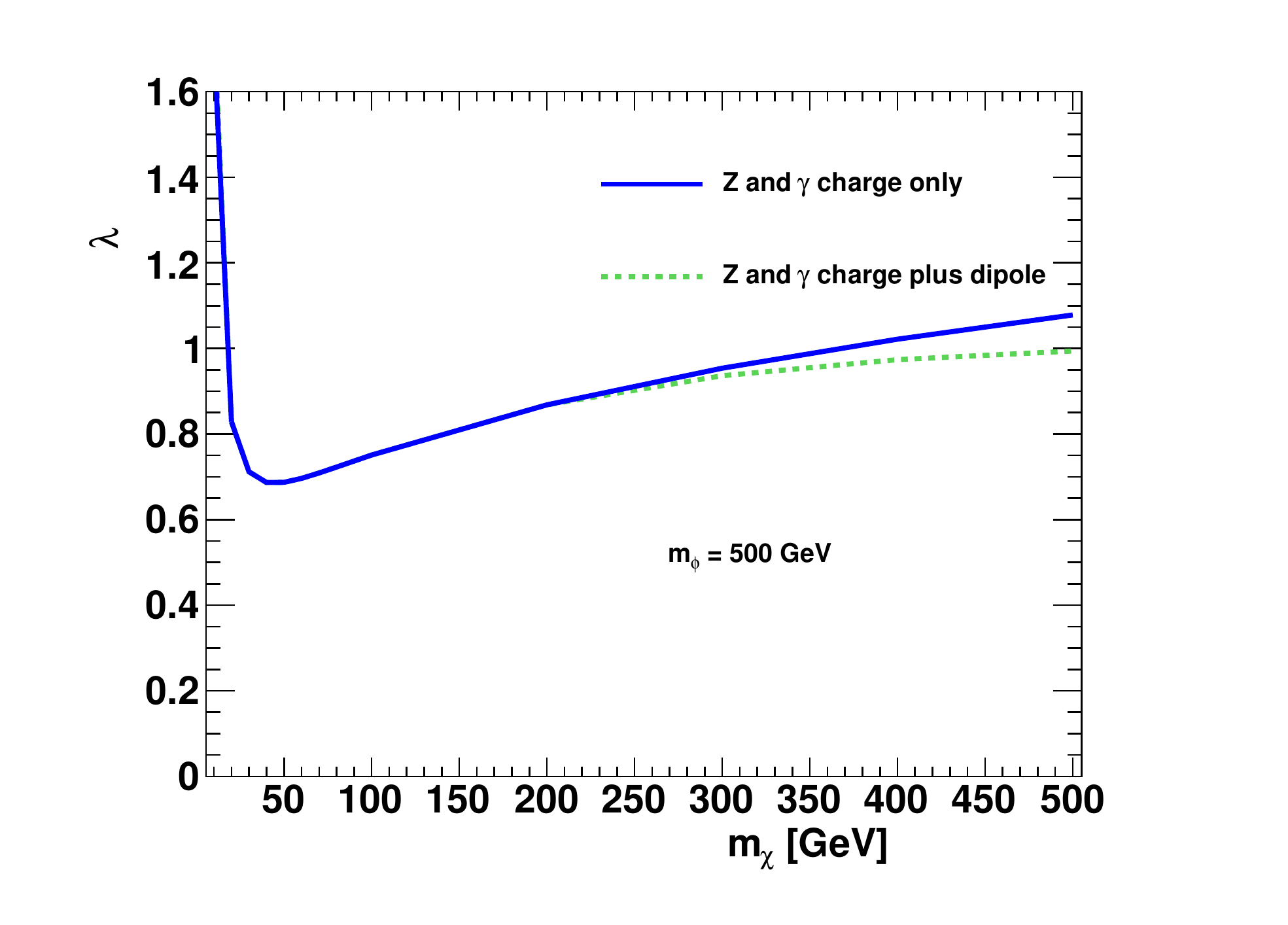} 
\caption{\small The LUX bound on the coupling $\lambda$ for $m_\phi = 500$ GeV calculated using the effective $Z$ coupling and charge term, with and without the addition of the dipole term. }
\label{fig:dddipole}
\end{center}
\end{figure}

So far we have kept the magnetic dipole terms. The momentum dependence in the magnetic dipole interaction makes it impossible to factor the differential event rate as a product of the elastic cross section and the velocity integration. Following the procedure in appendix C of Ref.~\cite{Hamze:2014wca}, we calculate the differential rate numerically, and work out the exclusion limits from LUX~\cite{Akerib:2013tjd} in the presence of the dipole terms. The result is shown in Fig.~\ref{fig:dddipole}. We find that the effect of the magnetic dipole operator is negligible compared to the effective $Z$ and charge operators in setting limits for the coupling $\lambda$. Based on this conclusion, for the rest of the paper we will neglect the magnetic dipole contributions, which will allow us to directly use the LUX bounds for the elastic scattering cross section in order to evaluate the constraints on the parameter space of the top FDM model.

\begin{figure}[tp]
\begin{center}
\includegraphics[width=0.55\textwidth]{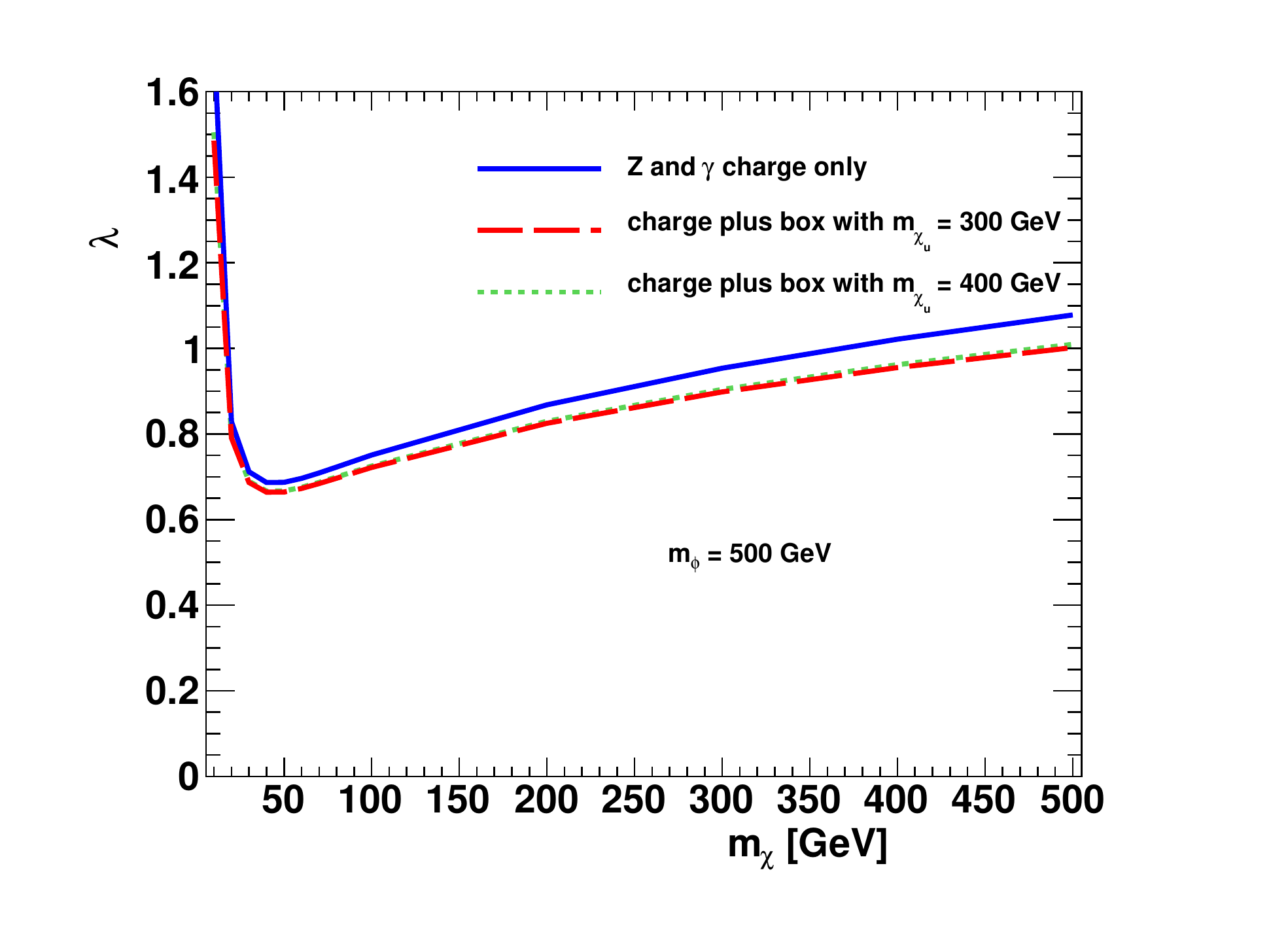} 
\caption{\small The LUX bound on the coupling $\lambda$ for $m_\phi = 500$ GeV calculated using the effective $Z$ coupling and charge term, with and without the addition of the effective coupling to $u$-$\bar{u}$. }
\label{fig:ddbox}
\end{center}
\end{figure}

In a similar fashion we study the importance of the effective four-point coupling to $u$-$\bar{u}$ by using the result of the box diagram calculated in Ref.~\cite{Kumar:2013hfa}. In Fig.~\ref{fig:ddbox} we plot our results in the form of the LUX bound on the coupling $\lambda$ with and without taking the box diagram into account, confirming the conclusions derived in Ref.~\cite{Kumar:2013hfa} that the direct detection bounds are dominated by the effective $\gamma$/$Z$ coupling. For the rest of this paper, we will use only this dominant contribution in calculating the bounds from direct detection experiments on the parameter space of the top FDM model.

\begin{figure}[htp]
\begin{center}
\includegraphics[width=0.45\textwidth]{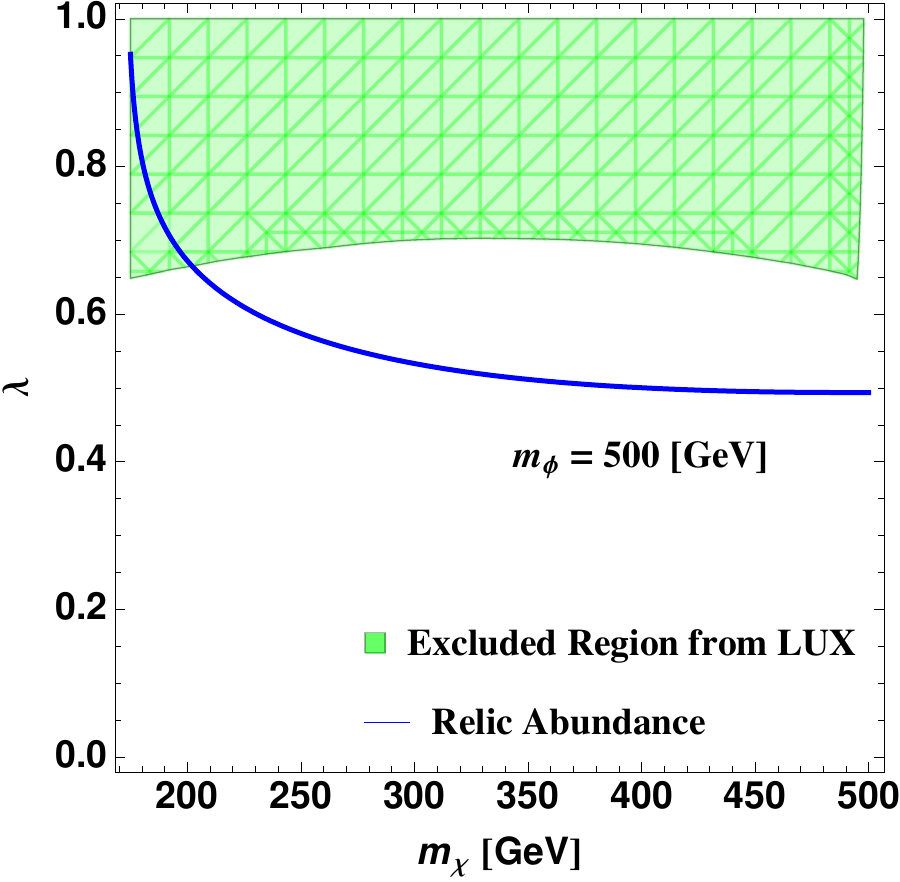} 
\includegraphics[width=0.45\textwidth]{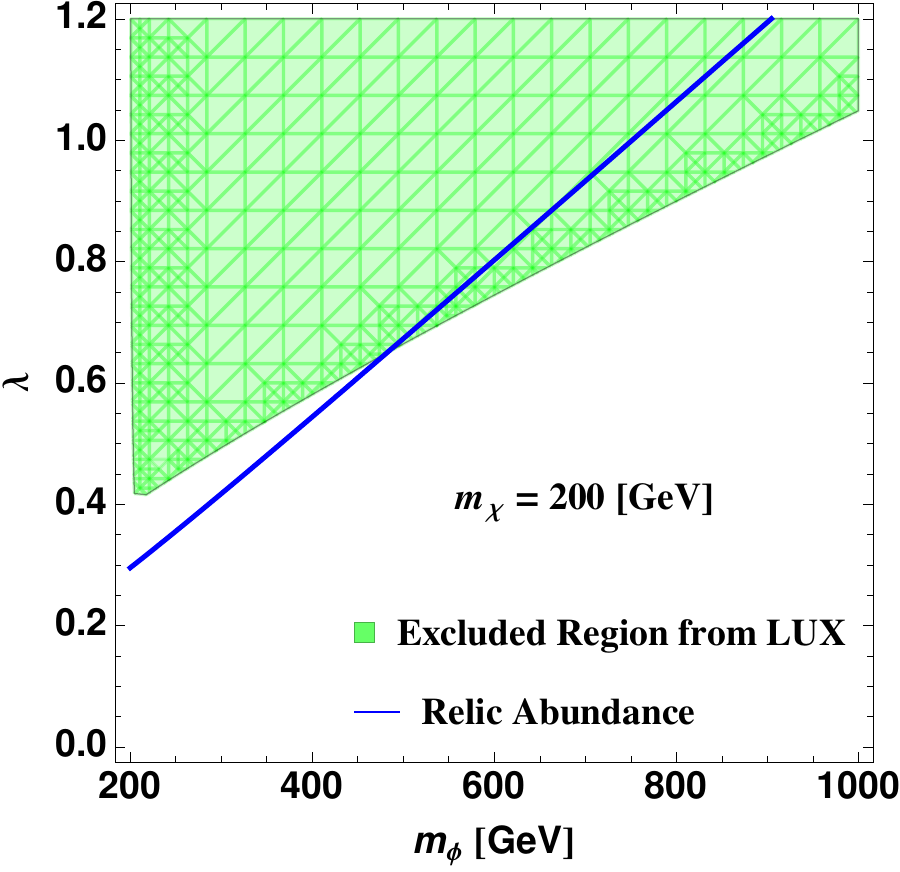}   
\caption{\small The LUX exclusion region in the  $\lambda$-$m_{\chi}$ plane ($\lambda$-$m_{\phi}$ plane) for $m_\phi=500~$GeV ($m_\chi=200~$GeV). The correct relic abundance is obtained on the blue curve.}
\label{fig:ddmasslambda}
\end{center}
\end{figure}

The leading (spin-independent) contribution to the matrix elements at the level of nucleons is
\begin{align}
	\langle \chi ,\, N \left| \overline{\chi} \gamma^\mu \chi \overline{N} \gamma_\mu N \right| \chi ,\, N \rangle &= 4 m_\chi m_N .	\label{eq:NmatrixF}
\end{align}
It is then convenient to define the dark matter-nucleon effective couplings
\bea
	{\mathcal C}^N &=&  4 m_\chi m_N c^N_Z  + 4 m_\chi m_N c^N_{\gamma},
	\label{eq:NeffF}
\eea
in terms of which the total scattering cross section (at the level of the nucleus) can be written as 
\begin{equation}
	\sigma_T  = \frac{1}{16 \pi} \left( \frac{1}{m_\chi + m_p} \right)^2 \left[ Z {\mathcal C}^p + (A-Z) {\mathcal C}^n \right]^2.
\label{eq:fdd}
\end{equation}

\begin{figure}[htp]
\begin{center}
\includegraphics[width=0.45\textwidth]{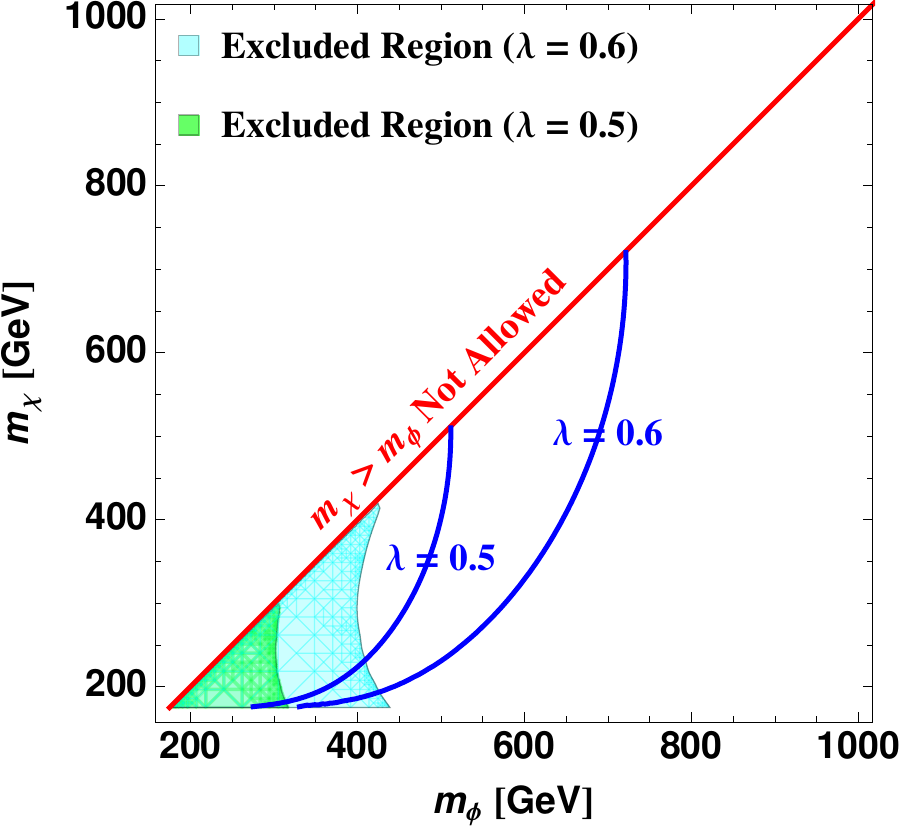} 
\caption{\small The LUX exclusion bounds in the $m_{\chi}$-$m_{\phi}$ plane for $\lambda = 0.5$ (green shaded region) and $\lambda=0.6$ (blue shaded region). For each case, we also show the curve on which the correct relic abundance is obtained.}
\label{fig:ddmassmass}
\end{center}
\end{figure}

We show in Fig.~\ref{fig:ddmasslambda} the LUX~\cite{Akerib:2013tjd} exclusion region in the $\lambda$-$m_{\chi}$ plane ($\lambda$-$m_{\phi}$ plane) for a generic value of $m_{\phi}$ ($m_{\chi}$). For $m_\phi=500~$GeV, $\lambda$ is constrained to be smaller than $0.7$. Note also that the correct relic abundance (shown by the blue curve) is obtained for $m_{\chi}\gsim200~$GeV. This is generic: Independent of the value of $m_{\phi}$, when $m_{\chi}$ gets close to the top mass, the relic abundance constraint favors large values of $\lambda$ which is in conflict with direct detection bounds. Also, for any fixed value of $m_\chi$, obtaining the correct relic abundance puts an upper limit on $m_{\phi}$ as can be seen in the second plot of Fig.~\ref{fig:ddmasslambda}.

\begin{figure}[htp]
\begin{center}
\includegraphics[width=0.45\textwidth]{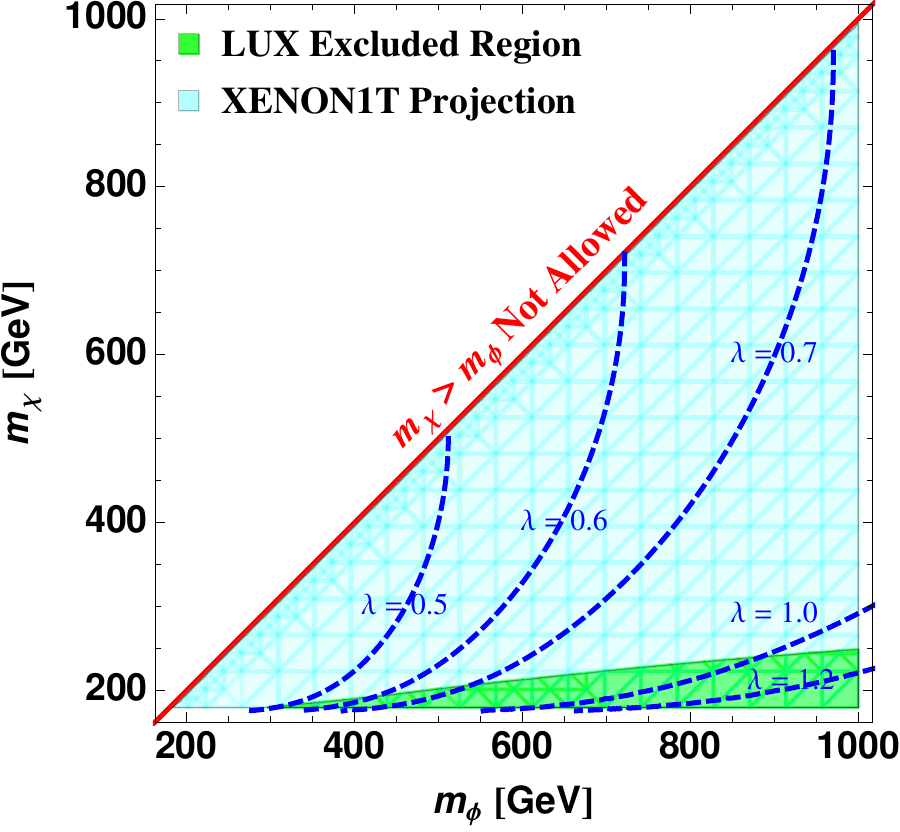} 
\caption{\small The current LUX exclusion bounds in the $m_{\chi}$-$m_{\phi}$ plane (green shaded region), where $\lambda$ is varied to give the correct relic abundance at each point. Contours of constant $\lambda$ are also shown for reference. The full parameter region in this plot (and beyond) can be excluded by the XENON1T experiment (blue shaded region).}
\label{fig:ddmassmass2}
\end{center}
\end{figure}

In Fig.~\ref{fig:ddmassmass} we show the LUX exclusion region in the $m_{\chi}$-$m_{\phi}$ plane for fixed values of $\lambda$. Not surprisingly, the exclusion region grows with $\lambda$. However, the curve on which the correct relic abundance is obtained also extends further, such that an order one fraction of the parameter space still remains viable. In order to illustrate this even more efficiently, in Fig.~\ref{fig:ddmassmass2}, we plot the (green shaded) LUX exclusion region by assigning to each point in the mass plane the value of $\lambda$ for which the correct relic abundance is obtained for that mass point. As can be seen from this plot, only relatively light $\chi$ masses are excluded. Contours corresponding to different values of $\lambda$ are also plotted for reference. 

Having mapped out the exclusion region from current constraints, we turn our attention to what region of parameter space can be probed with the next generation of direct detection experiments. We use the projected exclusion reach of the XENON1T experiment with an exposure of 2.2~ton years~\cite{XENON1T}, which is provided in the range $m_{\chi}\le1~$TeV. We find that top FDM (subject to relic abundance constraints) can be entirely excluded in this range. This is shown in Fig.~\ref{fig:ddmassmass2} as the blue shaded region. We also extrapolate the XENON1T projected exclusion reach to $m_{\chi}>1~$TeV by using $n_{\chi}\propto m_{\chi}^{-1}$ and assuming a constant detector efficiency for $m_{\chi}<10~$TeV. We find that the exclusion region extends to the full perturbative parameter range ($\lambda\le1$) for the model.

\section{Discovery Prospects at the LHC}
\label{sec:collider}

We now proceed to investigate the discovery prospects for top FDM at the LHC. We will focus on the case $m_{\chi}\gsim200~$GeV, which as we have shown in the previous section is strongly favored by direct detection experiments and relic abundance constraints. Searches based on kinematic features such as missing transverse energy ($\MET$) or transverse mass variables have not been very effective in probing this mass range at the end of the 7~TeV and 8~TeV runs of the LHC, and therefore we will focus our attention on discovery prospects at the 14~TeV run.

\subsection{Production and Decay Modes}

\begin{figure}[htp]
\includegraphics[scale=0.55]{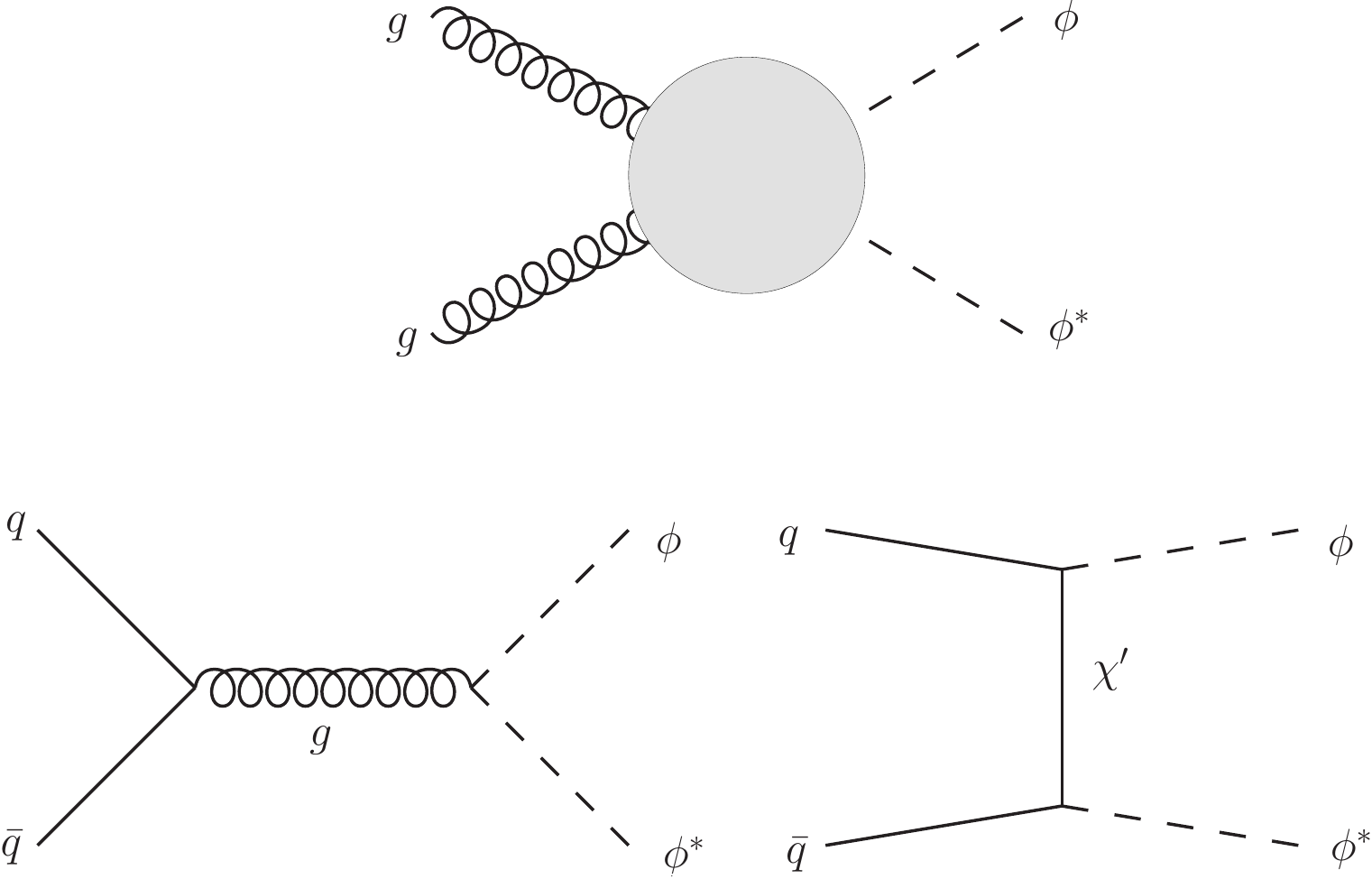}
\caption{Feynman diagrams for the pair production of the mediator $\phi$ at the LHC. }
\label{diagphiphi}
\end{figure}

\begin{figure}[htp]
\includegraphics[scale=0.3]{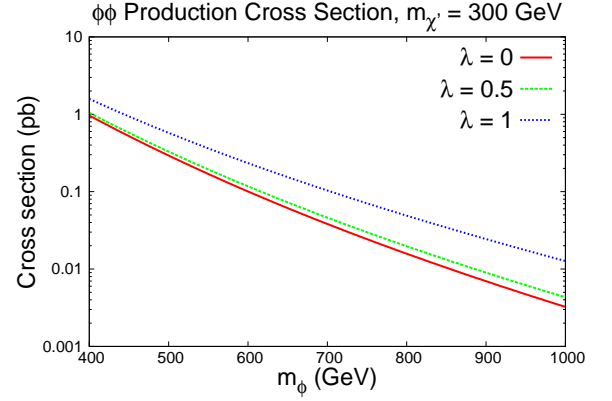}
\caption{$\phi$-pair production cross section for $\lambda = 0, 0.5, 1$ with $m_{\chi'}=300~$GeV.}
\label{xsecphiphi}
\end{figure}

Since $\chi$ is a SM singlet while $\phi$ is colored, the most important production mechanism is $\phi$-pair production from gluons in the initial state. This production mode is pure QCD and is therefore independent of $\lambda$. $\phi$ can also be pair produced through QCD interactions from a $q$-$\bar{q}$ initial state. However, in this case there is also a contribution from $t$-channel $\chi'$ (specifically, $\chi_{u}$) exchange, which has $\lambda$ dependence. Processes contributing to $\phi$-pair production are shown in Fig.~\ref{diagphiphi}, and the cross section for several choices of $\lambda$  are shown in Fig.~\ref{xsecphiphi}. We see that the $t$-channel $\chi'$ exchange can have an order one effect on the cross section for larger values of $\lambda$, which are not excluded by direct detection bounds.

\begin{figure}[htp]
\includegraphics[scale=0.55]{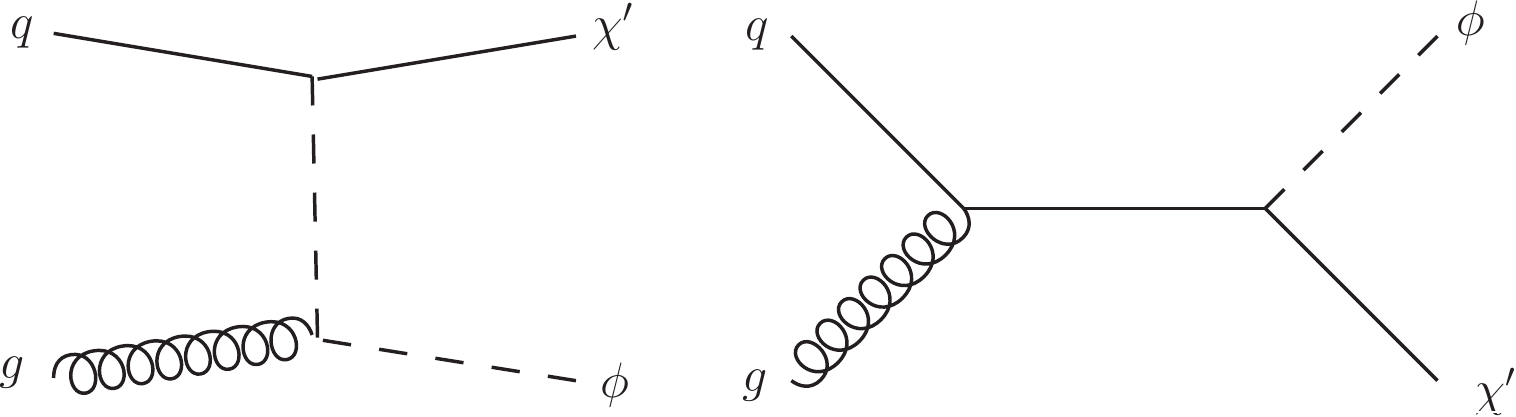}
\includegraphics[scale=0.55]{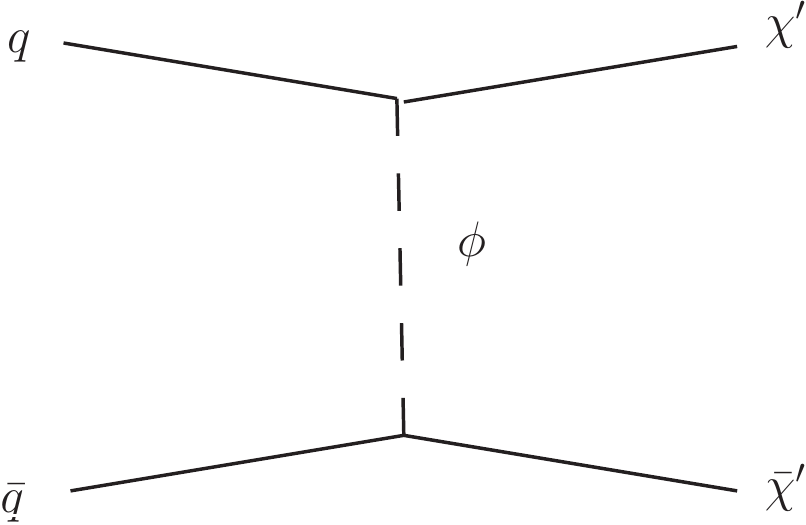}
\caption{Feynman diagrams for $\phi$-$\chi'$ associated production and $\chi'$-pair production at the LHC.}
\label{fig:diagchichi}
\end{figure}

\begin{figure}[htp]
\includegraphics[scale=0.3]{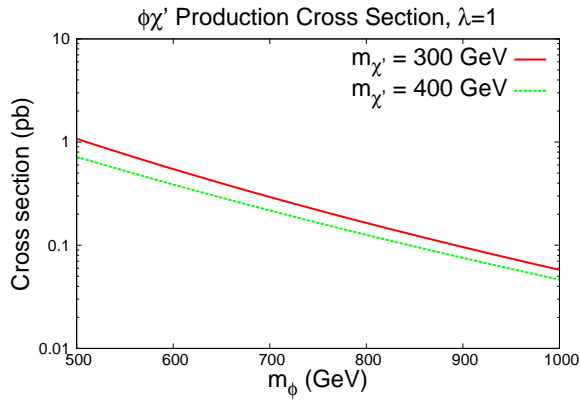}
\includegraphics[scale=0.3]{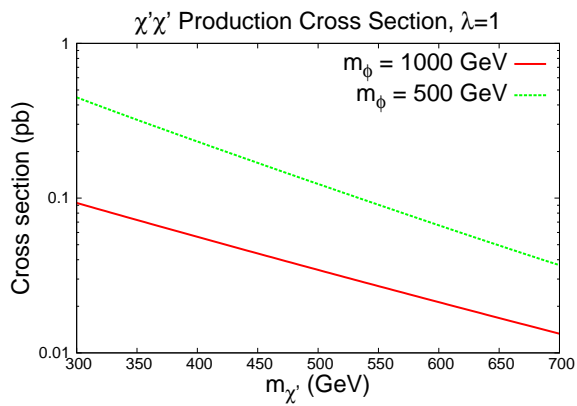}
\caption{(Top) $\phi$-$\chi'$ associated production cross section for $\lambda=1$ and two values of $m_{\chi'}$. $\sigma$ scales like $\lambda^{2}$. (Bottom) $\chi'$-pair production cross section for $\lambda=1$ and two values of $m_{\phi}$. $\sigma$ scales like $\lambda^{4}$.}
\label{fig:xsecphichi}
\end{figure}

$\phi$-pair production is not the only production channel in this model, since $\chi_{u}$ can couple to the initial state quarks through the FDM interaction of Eq.~\ref{eq:FDMterm}. In particular, both $\phi$-$\chi_{u}$ associated production and $\chi_{u}$-pair production are allowed, and these proceed via the Feynman diagrams of Fig.~\ref{fig:diagchichi}. Note that $\sigma(pp\to \phi \chi_{u})\propto \lambda^{2}$ while $\sigma(pp\to \bar{\chi}_{u} \chi_{u})\propto \lambda^{4}$. The cross sections for these channels are shown in Fig.~\ref{fig:xsecphichi}. 

\begin{figure}[htp]
\includegraphics[scale=0.7]{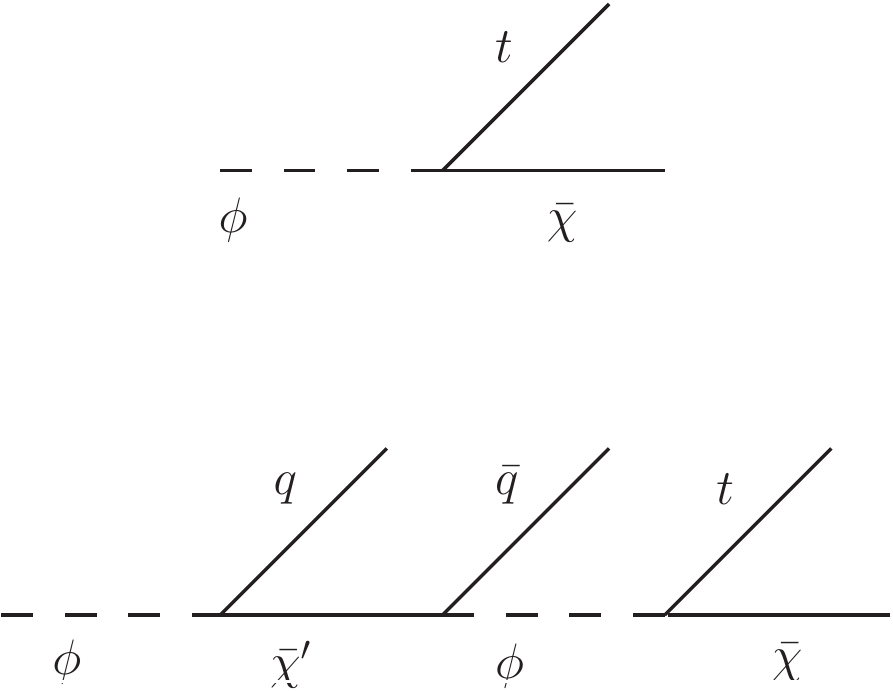}
\caption{The short and long decay modes for $\phi$.}
\label{fig:decays}
\end{figure}

\begin{figure}[htp]
\includegraphics[scale=0.32]{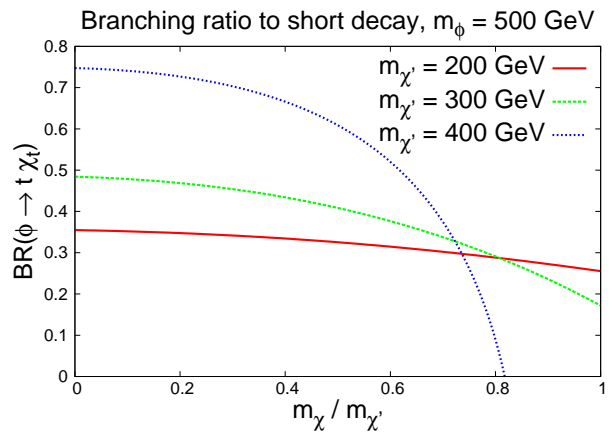}
\caption{Branching ratio of the short decay mode of $\phi$ as a function of the ratio $m_{\chi} / m_{\chi^\prime}$ for several values of $m_{\chi'}$. When $m_{\phi}-m_{\chi}$ drops below $m_{t}$, the short decay mode can only proceed through off-shell tops and therefore the branching ratio becomes negligible.}
\label{fig:BR}
\end{figure}

Note that all production channels involve $\phi$ and $\chi_{u}$, both of which decay promptly. Therefore both sides of each event will contain visible final states. $\phi$ has ``short'' and ``long'' decays, where in the former case it decays directly to a top quark and $\chi$, while in the latter case $\phi$ decays to a light quark and the corresponding flavor of $\chi^\prime$, which subsequently decays to a light quark, a top quark and a $\chi$ through an off-shell $\phi$. Depending on the mass difference between $\chi'$ and $\chi$, the top quark itself may be off-shell. This is illustrated in Fig.~\ref{fig:decays}. The branching ratio for the short decay mode is shown in Fig.~\ref{fig:BR} as a function of the ratio $m_{\chi} / m_{\chi^\prime}$ for several values of $m_{\chi^\prime}$. The branching ratio for the long decay mode (which is really two decay modes for $\chi_{u}$ and $\chi_{c}$, which have identical branching ratios to a very good approximation) can simply be found as one minus the branching ratio for the short decay mode. The production and decay modes under consideration ultimately result in a $t$-$\bar{t}$ pair and a $\chi$-$\bar{\chi}$ pair in every event (the tops possibly being off-shell depending on the details of the spectrum), and anywhere between zero and four additional light jets depending on the number of on-shell $\chi'$ in the event.

Note that while $m_{\chi}<m_{\phi}$ must be satisfied in order to obtain a neutral DM particle, if $\Delta m$ in Eq.~\ref{eq:FDMmass} is large enough, then $m_{\chi'}$ can exceed $m_{\phi}$. When that is the case, the collider phenomenology is particularly simple: all signal events basically arise from $\phi$-pair production followed by the short decay mode of $\phi$. The LHC discovery prospects of this simpler case are essentially the same as for the model studied in detail in Ref.~\cite{Gomez:2014lva}. Therefore in the remainder of this section we will focus our attention on how the collider prospects are affected by the presence of $\chi'$, and what kind of impact this may have on the conclusions reached in Ref.~\cite{Gomez:2014lva}. Note that the answer is not entirely obvious. On the one hand, one may argue that the presence of $\chi'$ may improve discovery prospects by opening new production modes, thus increasing the cross section, and by making the long-decay mode of $\phi$ possible, which gives rise to additional light jets in the event. On the other hand, for values of $\lambda$ that are allowed by direct detection bounds, the collider phenomenology is still dominated by $\phi$-pair production, thus the number of signal events does not significantly increase, and furthermore within these events, those with the now accessible long $\phi$ decays may be more difficult to distinguish from background, since the top quarks and the $\MET$ in these events are only produced in the last decay stage, which due to the presence of $\chi'$ has a smaller available phase space, therefore leading to softer tops and less $\MET$. Finally, the presence of $\chi'$ may allow for a more exotic signature in signal events, namely displaced vertices that occur in $\chi'$ decays, which may be more important than both of the above considerations for discovery. This last possibility however may or may not be realized depending on the mass spectrum.

In the following, we will first attempt to evaluate how the presence of $\chi'$ affects the discovery prospects for top FDM in a search based on transverse mass variables. We will do this by setting up a relatively simple analysis based on the dileptonic decay mode of the top quarks, which is the cleanest channel. We will look at several mass spectra for a fixed value of $m_{\phi}$ with varying $m_{\chi}$ and $m_{\chi'}$, since kinematic features of signal events will depend on the differences of $m_{\phi}$, $m_{\chi'}$ and $m_{\chi}$, but they will be less sensitive to the overall mass scale. In particular, as the mass differences in the spectrum change, top quarks produced in decays may be on or off shell, leading to very different efficiencies for analysis cuts. In contrast, the total signal normalization will depend mainly on the overall mass scale ($m_{\phi}$), but it will be insensitive to the mass differences.

After studying the impact of $\chi'$ in a kinematics-based search, we will return to the question of displaced vertices in $\chi'$ decays, and then comment on how these may allow all the parameters of the top FDM model to be determined experimentally.

\subsection{Search Based on Transverse Mass Variables}

{\bf Event Generation and Selection:} Since we will focus on the dileptonic decays of the top quarks, the signal events in our analysis are characterized by two $b$-jets, a pair of opposite sign leptons that are uncorrelated in flavor, and $\MET$. SM processes with similar final state particles constitute backgrounds for the top FDM signal. In order to model the relevant backgrounds, we generate samples of 500,000 events (2 million for $t$-$\bar{t}$) at 14 TeV center of mass energy for the backgrounds listed in Table~\ref{smbgtable}. The $t$-$\bar{t}$ background samples are generated using MC@NLO~\cite{Frixione:2002ik,Frixione:2003ei} whereas the remaining backgrounds are generated using MadGraph~\cite{Alwall:2014hca} at LO, and later rescaled with $K$-factors. The order of the $K$-factors, and the associated references are also listed in Table~\ref{smbgtable}.

\begin{table*}[t]
\begin{tabular}{cccc}
\hline 
Background & Order & $\sigma$ after cuts (pb) & Reference\\
\hline 

$t\bar t$ &NLO& 14.3 & \cite{smbg:tt} \\
$tW$ & NNLO &0.7 & \cite{smbg:tW} \\
$t\bar tZ$ &NLO& $2.3\times 10^{-3}$   & \cite{smbg:ttZ} \\

\hline 
\end{tabular}
\caption{Details of background simulation. The $t$-$\bar{t}$ background is generated at NLO, while the others are generated at LO and rescaled with $K$-factors. The order of each $K$-factor we use and the associated references are listed, as well as the cross section of each background after selection cuts.}
\label{smbgtable}
\end{table*}

We force all final state tops and $W$'s to decay leptonically, including leptonic decays of $\tau$'s. The $t\bar tZ$ background has a much smaller cross section compared to the others. However, we include it because our search strategy will be based on $\mt$~\cite{Lester:1999tx,Barr:2003rg,Burns:2008va} and $t\bar{t} Z$ it is the dominant background at high $\mt$. For this reason we force the $Z$ boson to decay to neutrinos, as any other decay mode does not contribute at high $\mt$ and is subdominant at low $\mt$.

In generating signal samples, we fix $m_{\phi}=500~$GeV. The production cross section plots of the previous subsection can be used to extrapolate the results of the analysis to different values of $m_{\phi}$. The mass points we use are given in the first two columns of Table~\ref{bestcut}. For each value of $m_{\chi}$ we also include a spectrum with $\chi'$ decoupled as a point of comparison in order to better assess the impact of the presence of $\chi'$ on discovery prospects. This is not unrealistic, since even for $m_{\chi'}$ moderately above $m_{\phi}$, $\phi$-pair production will be the dominant production mode and essentially no on-shell $\chi'$ will be produced. In order to be consistent with relic abundance and direct detection constraints, we only consider $m_{\chi}\ge200~$GeV, and we fix  $\lambda=1/2$. Note that this choice of $\lambda$ is conservative for the $\phi$-pair production cross section (see Fig.~\ref{xsecphiphi}), and decay kinematics are insensitive to $\lambda$. For each mass spectrum we generate 200,000 events at 14 TeV using MadGraph~\cite{Alwall:2014hca} for each of the signal production modes, which we combine with the appropriate ratios. All top quarks in signal samples are forced to decay leptonically.

Both signal and background samples are passed through FastJet~\cite{fastjet1,fastjet2} to reconstruct jets using the anti-$k_t$ jet clustering algorithm with radius $R = 0.4$~\cite{antikt}. We use a $b$-tagging efficiency of $0.6$. Isolated leptons are identified with cuts on the variable $p_T({\rm iso})$ around the lepton track, where the $p_T({\rm iso})$ is the sum over the $p_T$ of the tracks with $p_T > 1$ GeV for each track, within a $\Delta R < 0.2$ cone around the lepton candidate. For electrons, the isolation cut is chosen to be $p_T({\rm iso}) < 0.1 \times p_T({\rm lepton})$, while for muons, the isolation cut is $p_T({\rm iso}) < 1.8$ GeV. Any jet overlapping with an isolated lepton within $\Delta R < 0.2$ is eliminated from the jet candidates, and electron or muon candidates overlapping with any remaining jets within $\Delta R < 0.4$ are discarded to avoid accepting leptons from the decay of heavy hadrons.

Jet energy resolution is simulated using a Gaussian smearing with a variance $\sigma$:
\beq
	\frac{\sigma}{E} = \frac{S}{\sqrt{E}}\oplus  \frac{N}{E} \oplus C,
\eeq
where the terms $S, N, C$ are chosen to be $0.5$~GeV$^{1/2}$, $5$~GeV, and $0.04$, respectively. The $\eta$ and $\phi$ of the four-momentum of a jet is also smeared by a Gaussian with $\sigma=0.025$. Electron energies are smeared with $\sigma=0.02\times \sqrt{E/{\rm TeV}}$ and muon energies by $\sigma=0.1\times \sqrt{E/{\rm TeV}}$.

We also smear the transverse missing  energy of the $\MET$, defined as the magnitude of the two-vector momentum summed over all the final state invisible particles (neutrinos and LSPs). The individual $x$ and $y$ components of the $\MET$ vector are smeared by~\cite{METsmear} 
\beq
\sigma(\MET)_{x,y}[GeV]=(0.40+0.09\times\sqrt{\mu})\times\sqrt{H_{T}[GeV]+\mu\times 20},
\eeq
where $H_T$ is the scalar-summed $p_T$ of all visible final state particles within $|\eta| < 5.0$, and $\mu$ denotes the number of interactions per bunch crossing. For each event $\mu$ is chosen from a Gaussian distribution with central value equal to 50 and standard deviation equal to 10.

For the analysis, we use the following selection cuts in order to reduce backgrounds:
\begin{itemize}
\item At least two jets with $\pt > 20$~GeV and $|\eta| < 2.5$
\item  At least one $b$-tagged jet
\item Two opposite-sign leptons ($e$ and $\mu$ only) with $\pt > 20$~GeV and $|\eta| < 2.5$. Events with same flavor dileptons are vetoed if $|m_{\ell\ell}-m_{Z}|<10~$GeV
\item  $\MET>20~$GeV.
\end{itemize}
The cross section of each background sample after selection cuts is listed in Table~\ref{smbgtable}.

\begin{table*}[t!]
\begin{tabular}{cccccccc}
\hline 
$m_{\chi'}$~(GeV)	& $m_{\chi}$~(GeV)	& $\sigma_{\phi\phi}$~(fb)	& $\sigma_{\phi\chi}$~(fb)	& $\sigma_{\chi\chi}$~(fb)	& Best $\mt^\mathrm{cut}$~(GeV)	& $S/\sqrt B$ \\

\hline 

{\rm N/A} & 300 & 7.7 & - & - & & $< 1$ \\

400 & 300 & 9.0 & 3.4 & 0.11 & & $< 1$ \\

{\rm N/A} & 250 & 8.1 & - & - & 145 & 3.3 \\

400 & 250 & 9.0 & 4.8 & 0.29 & 135 & 2.0  \\

350 & 250 & 9.0 & 4.1 & 0.14 & 130 & 1.7  \\

{\rm N/A} & 200 & 8.4 & - & - & 165 & 8.9 \\

400 & 200 & 9.0 & 4.8 & 0.36 & 155 & 5.7  \\

350 & 200 & 9.0 & 5.3 & 0.39 & 145 & 3.5  \\

300 & 200 & 9.0 & 4.9 & 0.19 & 145 & 3.5  \\

\hline 
\end{tabular}
\caption{The signal spectra used in our analysis, with $m_{\phi}=500~$GeV. For each spectrum we list the cross section after selection cuts for each production mode, with $\lambda=0.5$. We also list the choice of the $\mt$ cut that maximizes $S/\sqrt B$, subject to the constraint that at least 50 signal events pass all cuts at 100~fb$^{-1}$, as well as the value of $S/\sqrt B$ for that choice. For the spectra with $m_{\chi}=300$~GeV, $S/\sqrt B$ remains below 1 for any choice of $\mt$ cut. For each value of $m_{\chi}$ we also include a spectrum where $\chi'$ is decoupled as a point of comparison in order to better assess the impact of the presence of $\chi'$ on discovery prospects.
}
\label{bestcut}

\end{table*}

{\bf Analysis and Results:} Our search strategy will rely on the kinematic variable $\mt$~\cite{Lester:1999tx,Barr:2003rg,Burns:2008va} in the leptons + $\MET$ part of the event, which is theoretically bounded from above by $m_{W}$ in events where the missing transverse energy arises entirely from two neutrinos originating from $W$ decays, whereas for events with additional sources of missing transverse energy, $\mt$ can be larger, and is therefore a good variable to separate signal from background. $\mt$ is defined as\footnote{In our analysis we only use $\mt$ with zero input mass for the invisible particle.} 
\bea
  \mt = \min_{\mathrm
\{  \,   \slashed{\mathbf p}_{T,1} +\slashed{\mathbf p}_{T,2} = \slashed{\mathbf p}_T   \} }
\left[ \max\left( m_T(\slashed{\mathbf p}_{T,1}, \ell_1), m_T(\slashed{\mathbf p}_{T,2}, \ell_2)\right)\right] \nn\\
\eea
where the minimization is performed over all possible partitions of the missing transverse energy into two parts, subject to $\slashed{\mathbf p}_{T,1} +\slashed{\mathbf p}_{T,2} = \slashed{\mathbf p}_T$. This models the neutrino transverse momenta in the event. $m_T(\slashed{\mathbf p}_{T,i}, \ell_i)$ is the transverse mass of the subsystem consisting of the hypothetical neutrino and the respective lepton, which for massless particles is defined as
\begin{equation}
m_T^2 ({\mathbf p}_{T,1},{\mathbf p}_{T,2}) =  2|{\mathbf p}_{T,1}| |{\mathbf p}_{T,2}|(1-\cos\Delta\varphi_{{\mathbf p}_{T,1},{\mathbf p}_{T,2}}).
\end{equation}

\begin{figure}
\includegraphics[scale=0.33]{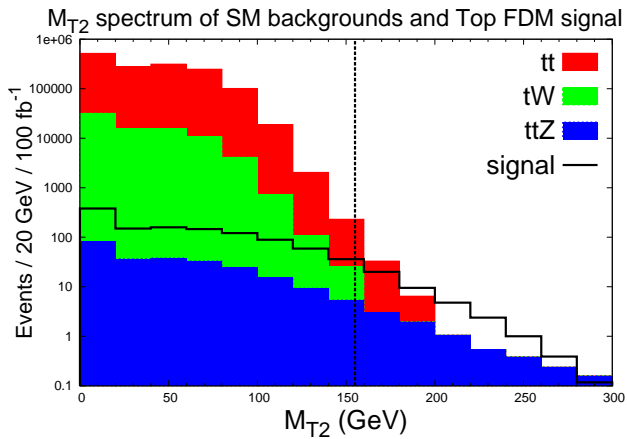}
\caption{$\mt$ histogram of background and the signal point $(m_\phi,m_{\chi^\prime},m_\chi)=(500,400,200)$~GeV with $\lambda=0.5$. The vertical line shows the best $\mt$ cut. Background histograms are stacked while the signal spectrum is shown separately.}
\label{MT2histogram}
\end{figure}

After the selection cuts described in the previous subsection have been applied, we look for the choice of an $\mt$ cut for each mass spectrum that will maximize $S/\sqrt B$, keeping the number of signal events after all cuts above 50 at 100 fb$^{-1}$ of integrated luminosity.  In Table~\ref{bestcut} we list the spectra we use, the cross sections for each production mode after selection cuts, and the optimal $\mt$ cut for each spectrum, along with the resulting value of $S/\sqrt B$. A plot of the $\mt$ spectrum for the background and a generic signal point ($m_{\chi^\prime} = 400$~GeV, $m_{\chi}=200$~GeV) is shown in Fig.~\ref{MT2histogram}.

These results are not difficult to interpret. The power of the $\mt$ based analysis arises from the production of $\chi$ particles with appreciable momenta. It is easy to see from Table~\ref{bestcut} that the most important factor for discovery prospects is the mass gap between $\phi$ and $\chi$, as this sets the upper limit of the $\chi$ momenta, in particular in the short decays of $\phi$ which yield the most energetic $\chi$ among all production and decay channels. We also see that the value of $m_{\chi'}$ is the second-most important quantity that impacts discovery prospects. The best cases in terms of $S/\sqrt{B}$ are obtained with the spectra in which $\chi'$ is decoupled. This shows that the significance of the search is driven by the short decay mode of $\phi$ which produces more energetic tops and more $\MET$. When $\chi'$ is reintroduced with a mass less than $m_{\phi}$, it affects the results of the analysis in two ways: 1) For fixed $m_{\phi}$ and $m_{\chi}$, as $m_{\chi'}$ increases, the branching ratio for the short decay mode increases, which boosts $S/\sqrt{B}$. 2) Once $m_{\chi'}$-$m_{\chi}$ goes above the top threshold, on-shell tops can be produced in the long decay mode of $\phi$ as well as in decays of $\chi'$ in the $\phi$-$\chi'$ and $\chi'$-$\chi'$ production modes. This boosts the signal significance as well.
 
\subsection{Displaced Vertices} 
\label{sec:displaced}

The lesson from the search based on transverse mass variables appears to be that the presence of $\chi'$ hurts rather than helps discovery prospects for top FDM. Thus the discovery reach mapped out in Ref.~\cite{Gomez:2014lva}, which corresponds to top FDM with $\chi'$ decoupled, is the best case scenario, while for $m_{\phi}>m_{\chi'}>m_{\chi}$, the events with on-shell $\chi'$ have less $\MET$ and less energetic tops. What on-shell $\chi'$ do add to signal events is light jets. However, it is difficult to find an optimal search strategy relying on light jets, both because light jets are also easily produced in the background from additional initial and final state emissions, and also because the $p_{T}$ spectrum of light jets in the signal will be highly dependent on the mass differences in the top FDM model and it is therefore not possible to devise cuts that will be optimal over the full parameter region.

\begin{figure}
\includegraphics[scale=0.33]{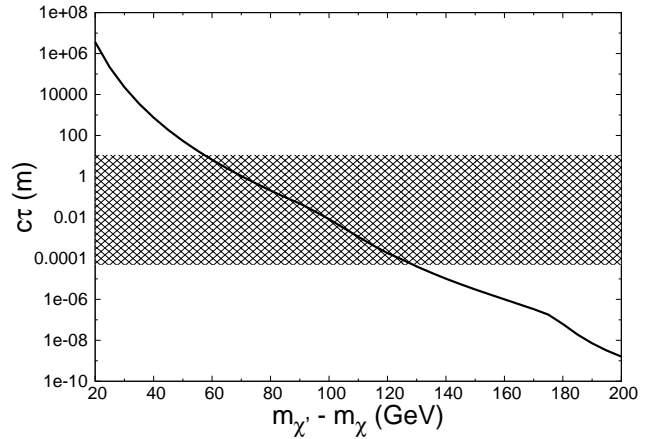}
\caption{The decay length $c\tau$ of $\chi'$ as a function of $m_{\chi'}$ for fixed values of $m_{\phi}=500~$GeV, $m_{\chi}=200~$GeV and $\lambda=0.5$. The results at very low mass splitting should be regarded as an idealization only, since in this regime deviations from MFV can cause flavor-changing $\chi'$ decays to become the dominant decay channel. The decay length changes by many orders of magnitude as $m_{\chi'}-m_{\chi}$ goes through the $m_{t}$ and $m_{W}+m_{b}$ thresholds. The range in which the LHC experiments are expected to be most sensitive to displaced decays is indicated as a gray-shaded horizontal band.}
\label{fig:displaced}
\end{figure}

There is however one case where on-shell $\chi'$, either directly produced via the $\phi$-$\chi'$ and $\chi'$-$\chi'$ production modes, or arising from the long decay mode of $\phi$, may significantly improve discovery prospects. When $m_{\chi'}-m_{\chi}\lsim m_{t}$, the $\chi'$ decay becomes four-body ($\chi'\to bWj\chi$). The lifetime of $\chi'$ is then in the range for displaced vertices. This is shown in Fig.~\ref{fig:displaced}. If $m_{\chi'}-m_{\chi}$ is even lower, namely below $m_{W}+m_{b}$, then even the $W$ boson must go off shell, and the $\chi'$ decay becomes five-body. While in the four-body regime the decay length $c\tau$ can be resolved with the silicon vertex detectors, in the five-body regime $c\tau$ becomes truly macroscopic. For at least a part of the five-body regime, one would expect to observe $\chi'$ decays occur in the calorimeters or even the muon system.

These conclusions assume that the FDM interaction of Eq.~\ref{eq:FDMterm} is diagonal in flavor space. However, we know that even if the UV theory respects MFV perfectly, there will be flavor-changing decay channels that $\chi'$ will inherit from quark mixing in the SM. This is in fact why we chose to include the term proportional to $\gamma$ in Eq.~\ref{eq:lambdamatrix}, as it allows us to estimate the size of the leading contribution to flavor-violating $\chi'$ decays in the exact MFV limit. Fortunately, even for $\gamma\sim{\mathcal O}(1)$, the flavor-violating (tree-level) three-body decay $\chi_{c}\to c\bar{c}\chi$ will be suppressed by $\left(y_{b}^{2}y_{c}V_{cb}\right)^{2}$ which makes it subdominant to the flavor-preserving decay even in the five-body regime. A flavor violating decay of $\chi_{u}$ would of course be even more suppressed, and therefore we conclude that the MFV-respecting corrections in Eq.~\ref{eq:lambdamatrix} do not affect our conclusions about displaced vertices. Loop-induced flavor-violating decays are even smaller, for instance $\chi_{c}\to\chi_{t}\gamma$ with $\phi$ and a charm quark running in the loop can be obtained from the above described tree-level flavor-violating decay by contracting the external charm quark lines and attaching a photon to the loop. The coupling structure in flavor space remains the same and therefore leads to the same suppression. Moreover, even though the decay is now two-body, the added suppression from the loop factor and $\alpha_{EM}\sim{\mathcal O}(10^{-2})$ makes the partial width for this decay channel even smaller than the tree-level flavor-violating decay.

In the absence of a complete UV model, we should of course only think of MFV as an approximation. However, deviations from MFV cannot be of order one in order to avoid tension with bounds from SM flavor-changing processes. Taking into account the suppression from the smallness of any off-diagonal $\phi$ couplings dictated by such SM constraints, the predictions for the four-body decay mode should be unaffected. On the other hand, whether or not the five-body decay mode can ever be dominant will depend on exactly how small such beyond-MFV off-diagonal couplings are. Therefore, when discussing the possibility of macroscopically displaced $\chi'$ decays, we should not forget that there is a significant amount of model dependence involved.

\begin{figure}
\includegraphics[scale=0.6]{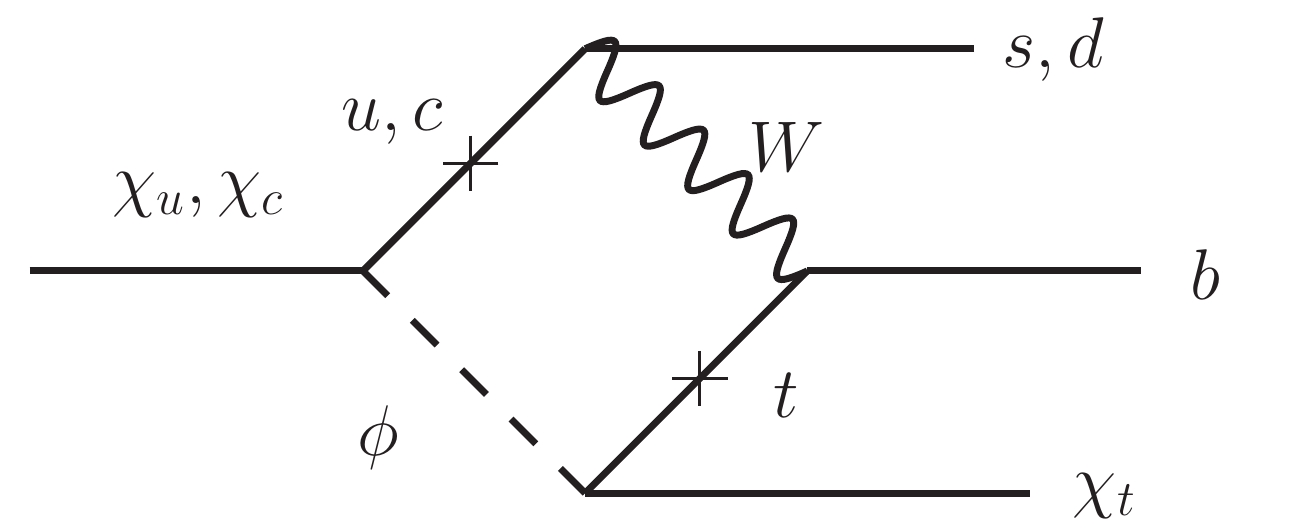}
\caption{The leading contribution to $\chi'$ decay which is flavor preserving, without a top quark in the final state.}
\label{fig:loopdecay}
\end{figure}

Note also that there are loop-induced processes that allow $\chi'$ to decay without any flavor violation even for very small $m_{\chi'}-m_{\chi}$. The leading such process is shown in Fig.~\ref{fig:loopdecay}. At first sight, this process may appear to be of the same size as the tree-level decay in the five-body region\footnote{Based on the very rough rule-of-thumb that the suppression from adding a loop is equivalent to the suppression from adding two final state particles at tree level.}. However, a closer inspection reveals this to be incorrect: both quark propagators in the loop are attached to an FDM vertex on one end and to a $W$ boson on the other. However, while the FDM interaction involves purely right-handed quarks, the weak interactions of course couple purely to left-handed quarks. Therefore both quark propagators in the loop require a mass insertion. This is not a big price to pay for the top quark, however already for the charm quark, this mass insertion ensures that the loop induced decay remains subdominant to the tree-level decay, even in the five-body regime.

\subsection{Measuring the Parameters of the Model}
\label{sec:params}

In the rest of this section we will comment on how the parameters of the top FDM model can be measured once discovery level significance has been achieved. As our conclusions so far have made clear, the answer to this question will be very different in various regions of parameter space of the model. Below, we will go through the distinct possibilities, starting from the case where certain parameters simply cannot be determined, towards more optimistic cases where all model parameters can be measured. Note that while the procedures we will outline below can work {\it in principle}, they will likely require a high-statistics signal sample, and whether or not this condition can be met will depend on the details of the model, in particular the $\phi$ mass which is the primary factor in setting the overall production cross section. The full list of model parameters is the mediator mass $m_{\phi}$, the DM multiplet masses $m_{\chi}$ and $m_{\chi'}$ (or equivalently $m_{0}$ and $\Delta m$ of Eq.~\ref{eq:FDMmass}), and the couplings $\lambda_{t}$ and $\lambda_{u,c}$ which may in principle be different (or equivalently $\alpha$ and $\beta$ of Eq.~\ref{eq:lambdamatrix}).

{\bf Case 1) $\chi'$ decoupled:} As discussed earlier, if $m_{\chi'}>m_{\phi}$, essentially no $\chi'$ are produced on shell. The LHC phenomenology is dominated by $\phi$-pair production and furthermore for the values of $\lambda_{t}$ that are allowed by direct detection and relic abundance constraints, the production in this channel is dominated by QCD processes (see Fig.~\ref{xsecphiphi}). This means that experimentally $m_{\chi'}$ and $\lambda_{u,c}$ play no role anywhere, and therefore cannot be measured.

Since we are considering measurements that will be performed once discovery level significance has been achieved, the kinematic information present in the data can be used for measuring the mass difference $m_{\phi}-m_{\chi}$, such as the $\MET$ spectrum or the endpoint of the $\mt$ distribution of the event including the $b$-jets (rather than the $\mt$ observable we used in the analysis presented earlier in this section which was constructed from the leptons and $\MET$ alone). Combined with the signal production cross section, which is a function of $m_{\phi}$ alone, the two masses $m_{\phi}$ and $m_{\chi}$ can then be independently determined. The parameter $\lambda_{t}$ does not affect the collider phenomenology. However, once the $\phi$ and $\chi$ masses have been measured, one can make a prediction for the value of $\lambda_{t}$ based on the relic abundance constraint. Furthermore, in the event that a signal is observed in a direct detection experiment, the predicted value of $\lambda_{t}$ can be cross-checked with the signal event rate in that experiment.

{\bf Case 2) $m_{\chi'}<m_{\phi}$, $m_{\chi'}-m_{\chi}\gsim m_{t}$:} Signal events in this case will sometimes contain on-shell $\chi'$, either directly produced, or originating from the long decays of $\phi$. However, the mass splitting between $\chi'$ and $\chi$ is large enough that $\chi'$ decays are three-body and prompt. The measurements discussed in case 1 can still be performed\footnote{Note that even though several production channels may be contributing to the signal, the events with the highest $\mt$ still come from the short decay mode of $\phi$, so the $\mt$ endpoint yields the same information as in case 1.}. 

In order to go any further, one needs a robust way to separate out signal events with and without on-shell $\chi'$. Since there are no displaced vertices, and since top quarks produced in $\chi'$ decays are on-shell, there is no simple way of doing so based on kinematics, although one may rely on the existence of additional light jets in the event which occur in events containing at least one $\chi'$. If requiring additional hard jets allows one to extract a signal subsample containing on-shell $\chi'$, then a detailed kinematic study of $\chi'$ decays, combined with a knowledge of $m_{\phi}$ and $m_{\chi}$ determined as in case 1 can then be used to infer the $\chi'$ mass.

As long as $\chi'$ decays are prompt, kinematics alone will not yield any information on the final remaining parameter $\lambda_{u,c}$. However, this coupling does have phenomenological importance: The production cross section for the $\phi$-$\chi'$ channel scales as $\lambda_{u,c}^{2}$, the cross section for the $\chi'$-$\chi'$ channel scales as $\lambda_{u,c}^{4}$, and the short vs. long branching ratios of $\phi$ depend on $\lambda_{u,c}$ as well\footnote{In principle even the production cross section in the $\phi$-$\phi$ channel has a nontrivial $\lambda_{u,c}$ dependence, however if $\lambda_{u,c}\lsim 0.5$, this additional contribution is subdominant and therefore difficult to observe.}. One can therefore attempt to measure $\lambda_{u,c}$ by separating the signal into bins with different numbers of light jets, and performing a global fit to the number of signal events in each bin using the predicted production cross sections and branching ratios in all channels as a function of $\lambda_{u,c}$.

\begin{figure}
\includegraphics[scale=0.33]{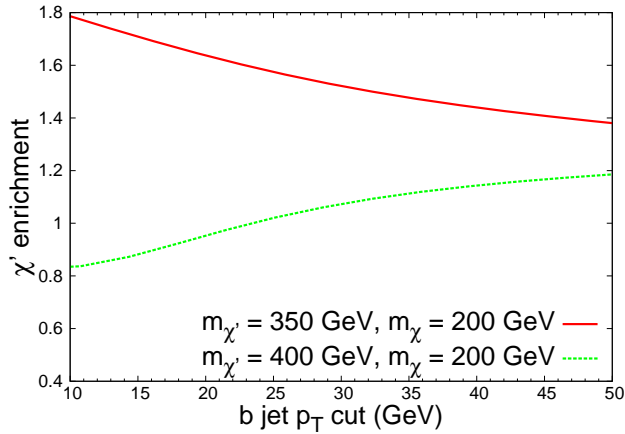}
\caption{The $\chi'$ enrichment factor, defined as the enhancement of the fraction of signal events containing at least one $\chi'$ after demanding that both $b$-tagged jets in the event have $p_{T}<X$, with $X$ plotted on the x-axis. The red curve is obtained from a signal spectrum with $m_{\chi'}-m_{\chi} < m_{t}$ while the green curve is obtained from a signal spectrum with $m_{\chi'}-m_{\chi} > m_{t}$. As described in the main text, the $\chi'$ content of the former signal point can be enhanced using such a cut with a low $X$ because $\chi'$ decays with off-shell tops give rise to softer $b$-jets, whereas for the latter signal point with on-shell tops, the $b$-jets tend to be harder.}
\label{fig:bjetcut}
\end{figure}

{\bf Case 3) $m_{\chi'}<m_{\phi}$, $m_{\chi'}-m_{\chi}\lsim m_{t}$ but no displaced vertices:} Note that the $\chi'$ decays do not become displaced immediately as 
the $\chi'$-$\chi$ mass splitting drops below the top quark mass. In fact, depending on the value of $\lambda$, the $\chi'$ decay length only becomes large enough to be detected by the LHC experiments when the mass splitting is 120--130~GeV (see Fig.~\ref{fig:displaced}). However, even though displaced vertices are not observable, there is a method based on kinematics in this case that allows one to obtain a subsample with a higher fraction of events containing on-shell $\chi'$. This relies on the following observation originally made in Ref.~\cite{Agashe}: the $p_{T}$ spectrum of $b$-jets produced in on-shell top quark decays peaks at a value that is independent of the boost of the original top quark. When the top quark is off-shell however, the $b$-jet $p_{T}$ spectrum is peaked towards zero. This suggests that by imposing a suitably chosen {\it upper} cut on $b$-jet $p_{T}$'s, one can reject more events with no on-shell $\chi'$ than those with on-shell $\chi'$. This is illustrated in Fig.~\ref{fig:bjetcut}. After obtaining a $\chi'$-enriched subsample of events this way, one can then attempt to measure $m_{\chi'}$ and $\lambda_{u,c}$ in a similar way as described for case 2.

{\bf Case 4) $m_{\chi'}<m_{\phi}$, with displaced vertices:} When the mass splitting between $\chi'$ and $\chi$ is less than 120--130~GeV, the $\chi'$ decays contain displaced vertices. This makes it much easier to obtain a signal subsample with on-shell $\chi'$, and determine $m_{\chi'}$ by studying the kinematics of $\chi'$ decays. Furthermore, one now has a much more robust handle on $\lambda_{u,c}$, since it can be measured directly from the decay length of $\chi'$.


\section{Conclusions}
\label{sec:conclusions}

We have studied the phenomenology of top FDM at direct detection experiments and at the LHC. In order to obtain the correct relic abundance, the preferred mass of the DM particle should be 200~GeV or higher. This is entirely compatible with current bounds from direct detection experiments. We have also studied the prospects of next generation direct detection experiments to look for top FDM and we have shown that the full perturbative parameter region ($\lambda\le1$) will be probed by the XENON1T experiment with 2.2~ton years of exposure.

There are several production and decay channels for top FDM at the LHC which populate final states with two (possibly off-shell) top quarks, missing transverse energy and possibly additional light jets. In the case that the heavier DM flavors $\chi'$ are not readily produced at the LHC, the collider phenomenology is simple, and similar to that of other models that have been studied previously in the literature such as the model of Ref.~\cite{Gomez:2014lva}. We have therefore focused mainly on the differences with this simpler case when $\chi'$ does play a role in the phenomenology. In particular, we have shown that when the DM flavors are sufficiently split in mass, the presence of $\chi'$ actually reduces the discovery reach of searches based on transverse mass variables. However, when the mass splitting is less than 120--130~GeV, we have shown that the decay length of $\chi'$ will be in a range that leads to displaced vertices. In this case, signal events are much easier to separate from background and the presence of $\chi'$ significantly improves discovery prospects.

We have also commented on how the parameters of the underlying model can be experimentally determined. Even in the case when $\chi'$ are not produced on shell, a combination of kinematic variables, the signal cross section and relic abundance constraints should allow one to determine the mass of the DM particle, the mediator particle as well as their coupling to the top quark, which can be cross-checked with results from future direct detection experiments. When $\chi'$ is readily produced on shell, the remaining model parameters, namely the mass of $\chi'$ and its coupling to the lighter quarks, also become experimentally observable. When $\chi'$ decays result in on-shell tops, this is only possible by relying on the properties of hard light jets in the event, which is not a simple measurement due to the limited signal statistics and to the inherent systematical uncertainties in making predictions concerning detailed properties of additional jets using Monte Carlo methods. When $\chi'$ decays are prompt but they result in off-shell tops, it is kinematically possible to enrich the $\chi'$ content of the data using a veto on hard $b$-jets, which makes it easier to attempt measuring these additional parameters of the model. In the case where $\chi'$ decays give rise to displaced vertices however, the determination of these additional parameters becomes much more straightforward. The presence of displaced vertices allows one both to obtain a sample of events containing on-shell $\chi'$ which can subsequently be studied using kinematic observables, and to directly measure the relevant coupling from the decay length.


\section*{Acknowledgements}

We thank Prateek Agrawal, Zackaria Chacko and Kathryn Zurek for helpful discussions and valuable comments. The research of the authors is supported by the National Science Foundation under Grant Numbers PHY-1315983 and PHY-1316033.


\end{document}